\documentstyle[sprocl]{article}

\def\Journal#1#2#3#4{{#1} {\bf #2}, #3 (#4)}

\def\NPB{{\em Nucl. Phys.} B}
\def\PLB{{\em Phys. Lett.}  B}

\def\PRD{{\em Phys. Rev.} D}
\def\RPP{\em Rept. Prog. Phys.}

\def\ARNPS{\em Ann. Rev. Nucl. Part. Sci.}
\def\PU{\em Phys. Usp.}

\def\simlt{\stackrel{<}{{}_\sim}}

\def\be{\begin{equation}}
\def\ee{\end{equation}}
\def\bea{\begin{eqnarray}}
\def\eea{\end{eqnarray}}

\begin{document}
\begin{titlepage}
\title{
ELECTROWEAK BARYOGENESIS WITH COSMIC STRINGS?}
\vspace{1cm}
\author{ J.R. ESPINOSA \\
CERN TH-Division\\
CH-1211 Geneva 23, Switzerland}

\maketitle
\vspace{1.cm}
\def\baselinestretch{1.15}
\begin{abstract}
I report on a critical analysis of the scenario of
electroweak
baryogenesis mediated by nonsuperconducting cosmic strings.
This
mechanism relies upon electroweak symmetry restoration in a
region around
cosmic strings, where sphalerons would be unsuppressed. I
discuss the
various problems this scenario has to face, presenting a
careful
computation of the sphaleron rates inside the strings, of
the chemical
potential for chiral number and of the efficiency of
baryogenesis in
different regimes of string networks. The conclusion is that 
the
asymmetry in baryon number generated by this scenario is
smaller than the
observed value by at least 10 orders of magnitude.\end{abstract}
\vspace{4cm}
\leftline{CERN-TH/99-05}
\leftline{January 1999}

\thispagestyle{empty}

\vskip-15.cm
\rightline{{ CERN-TH/99-05}}
\rightline{{ hep-ph/9901310}}

\vskip3in

\end{titlepage}

\title{
ELECTROWEAK BARYOGENESIS WITH COSMIC STRINGS?
}

\author{ J.R. ESPINOSA }

\address{
CERN TH-Division\\
CH-1211 Geneva 23, Switzerland}
 
\maketitle\abstracts{
I report on a critical analysis of the scenario of electroweak
baryogenesis mediated by nonsuperconducting cosmic strings. This
mechanism relies upon electroweak symmetry restoration in a region around
cosmic strings, where sphalerons would be unsuppressed. I discuss the
various problems this scenario has to face, presenting a careful
computation of the sphaleron rates inside the strings, of the chemical
potential for chiral number and of the efficiency of baryogenesis in
different regimes of string networks. The conclusion is that the
asymmetry in baryon number generated by this scenario is smaller than the
observed value by at least 10 orders of magnitude.}

\section{Introduction}

Electroweak baryogenesis \cite{review} is a beautiful idea which fails
(or is about to fail) in the best motivated models we have for physics
at the Fermi scale ($\sim 100\ GeV$). In the Standard Model, LEP II
experiments set a lower 
bound on the mass of the Higgs boson of about 97 GeV, implying that the
electroweak phase transition in that model is not first order but rather
a crossover \cite{SM}. In the Minimal Supersymmetric Standard Model the
electroweak
phase transition can be first order and sufficiently strong to allow for
electroweak baryogenesis, but this occurs in a very small region of
parameter space \cite{MSSM} which presumably will be ruled out by
LEP II in a couple of years.

One may take the previous negative results as
indication that the asymmetry in baryon number was not created at the
electroweak epoch, but rather related to the physics of $B-L$ violation
and neutrino masses. To stick to electroweak baryogenesis one can
consider extensions of the particle
content of the model to get a stronger electroweak phase transition (e.g.
extensions which include singlets). In this talk I will consider 
another possibility: how the remnants of physics at energy scales higher
than the electroweak  scale (cosmic strings in this case) can be useful
to overcome the problems of having a weak electroweak phase transition.

Electroweak baryogenesis requires the co-existence of regions of large
and small $\langle \varphi \rangle/T$, where $T$ is the temperature and 
$\langle \varphi \rangle$ the ($T$-dependent) Higgs vacuum expectation
value. At small or zero $\langle \varphi \rangle/T$ sphalerons are
unsuppressed and mediate baryon number violation, while large $\langle
\varphi
\rangle/T$ is needed to store the created baryon number (for $\langle
\varphi \rangle/T\geq 1$ sphaleron transitions are ineffective and
baryon number is conserved). Below the critical temperature $T_c^{EW}$
of the electroweak phase transition and irrespective of 
whether it is first or second order, $\langle \varphi \rangle/T$
grows until sphaleron transitions are shut-off. For baryogenesis to be
possible at those times, we need some region where $\langle \varphi
\rangle$ is forced to remain zero or small. The idea we examine in this
talk is that this can be the case along topological defects (like
cosmic strings) left over from some other cosmological phase transition
that took place before the electroweak epoch \cite{brand}.
If the electroweak symmetry is restored in some region around the
strings, sphalerons could be unsuppressed in the string cores while
they would be ineffective in the bulk of space, away from the strings.
The
motion of the string network, in a similar way as the motion of bubble
walls in the usual first-order phase-transition scenario, will leave a
trail of net baryon number behind.

Some problems with this scenario come immediately to mind. First, it is
clear that the space swept by the defects is much smaller than the total
volume, so there will be a geometrical suppression factor with respect to
the usual bubble-mediated scenario \cite{brand}.
Another suppression factor arises from the fact that there is a
partial cancellation between front and back walls of the string, which
tend to produce asymmetries of opposite signs \cite{brand}.
Another problem comes from the condition that the symmetry restoration
region (which naively would be of size $R_{rest}\sim
1/\sqrt{\lambda}\langle\varphi\rangle$, where $\lambda$ is the quartic
Higgs
coupling) should be large enough to contain sphalerons (which in the
symmetric phase have size $R_{sph}\sim 1/g^2T$), while outside the
strings, sphalerons should be suppressed ($\langle\varphi\rangle/T\geq
1$).
Combining both conditions one obtains $\lambda\leq g^4$, which means the
scenario would require small values of the Higgs mass, in conflict
with experimental bounds. LEP II tells us that $\lambda$ is at
least of order $g^2$, so that sphalerons won't fit in the restoration
region. In other words, for realistic values of the Higgs mass 
sphalerons are not going to be fully unsuppressed.
We will measure how effective they are by writing the rate of
sphaleron transitions per unit time and unit of string length as
$\Gamma_l=\kappa_l \alpha_w^2 T^2$. For a string with
$R_{rest}=R_{sph}$, one has $\Gamma_l R_{rest}^2$ equal to the rate in
the symmetric phase, corresponding to $\kappa_l\sim 1$. Values of
$\kappa_l$ much smaller than 1 would mean that sphalerons are not really
unsuppressed inside the strings.

In the rest of the talk I review the careful analysis of this
mechanism contained in ref. \cite{us}, to which I refer the
interested reader for further details.

\section{Strings with electroweak symmetry restoration}

Cosmic strings \cite{cosmics} are 1-dimensional solitons, stable by
topological reasons,
that can form in the spontaneous breaking of a symmetry $G$ where
I consider  the simplest case, $G=U(1)$, in this talk.
A model with a complex scalar $S$ and lagrangian
\be
{\cal L}=\partial_\mu S^*\partial^\mu S - \lambda_S(S^*S-S_0^2)^2,
\ee
admits global strings: configurations with $S=0$ along some line
(say the $z$-axis) and $S(r)=f(r)S_0 e^{i\theta}$, with
$f(\infty)\rightarrow 
1$, where $r$ is the distance to the $z$-axis and $\theta$ the azimuthal
angle. The radius of these strings (where most of the energy is trapped)
is set by the scale $1/m_S\equiv 1/\sqrt{\lambda_S}S_0$.

If the $U(1)$ is made local, in addition to the $S$ field, a non-zero
gauge field is also present, $A_{\mu}=-a(r)\partial_{\mu}\theta/q_S$,
with $a(\infty)=1$, where $q_S$ is the $U(1)$ charge of the $S$ field. 
This gauge field
is such that the covariant derivative $D_\mu S$ goes to zero for large
$r$ resulting in a finite energy per unit length of string.

We assume that $S$-strings (global or local) form at some temperature
$T_c^S>T_c^{EW}$ and are present at the time of the electroweak phase
transition. To force $\langle\varphi\rangle\rightarrow 0$ in the
cores of the strings, the Higgs field must interact either with the $S$
field or the $A_\mu$ field (if the strings are local):

\subsection{$S-\varphi$ interaction}

Suppose the scalar potential has the form
\be
V(S,\varphi)=\lambda_S(|S|^2-S_0^2)^2-\gamma(|S|^2-S_0^2)(|\varphi|^2-
\varphi_0^2)+ \lambda(|\varphi|^2- \varphi_0^2)^2,
\ee
with $\gamma>0$. The mass squared of the Higgs field in the string
background is
$m_\varphi^2(r)\sim\gamma(S_0^2-|S(r)|^2)-2\lambda\varphi_0^2$,
which is negative outside the string core but can be positive inside,
so that electroweak symmetry tends to be restored along the strings.
Exploring the ($S_0,\lambda_S,\gamma,\lambda$) parameter space, the
typical case, with $\lambda_SS_0^2\gg \lambda\varphi_0^2$ leads to
$R_{rest}\sim 1/m_\varphi(\infty)$. The best posible case to get a large
restoration region has $\lambda_S\ll \gamma\ll \lambda$ and $S_0\gg
\varphi_0$ and gives
$R_{rest}\sim\sqrt{\gamma/\lambda_S}/m_\varphi(\infty)$.

\subsection{$S-A_\mu$ interaction}

In this case we assume that the Higgs field carries a charge $q_\varphi$ 
under the extra $U(1)$ responsible for the strings, so that its
covariant derivative has an extra piece. As we saw, the $A_\mu$ field 
in the string goes like $-1/q_Sr$ at large $r$ to cancel the
azimuthal derivative of $S$, give vanishing
$D_\mu S$ and minimize energy. In $D_\mu \varphi$,  
the $A_\mu$ contribution is now proportional to $q_\varphi/q_S$
and the azimuthal derivative of $\varphi$ can cancel $D_\mu \varphi$
only if $q_\varphi/q_S$ is an integer. 
If that is not the case, a $Z_\mu$ boson condensate is induced until the
covariant derivative is cancelled \cite{davis}. 
In any case, a non-zero winding of $\varphi$ forces
$\varphi\rightarrow 0$ in the string core ($r=0$).
The restoration region around $r=0$ is larger in the presence of a
non-zero $Z_\mu$
string (case of non-integer $q_\varphi/q_S$).

\section{Sphaleron rates and CP asymmetry in the string cores}

In general, with no tuning of potential parameters nor a $Z_\mu$ 
condensate,
$\langle\varphi\rangle$ is zero only at the string core ($r=0$) and rises
inmediately away from that line. As the symmetry is never really restored 
in a wide region, the energy of the sphaleron
in such background (it can be computed in the lattice looking for a
saddle point of the energy functional) is only about a factor 0.7 smaller
than the sphaleron energy in the broken phase (alternatively
$\kappa_l\sim 10^{-6}$: that
is, sphalerons are not really unsuppressed in this type of strings).

The situation is better when a $Z_\mu$-field is induced, in which case
$\kappa_l\sim 1/30$ for $\langle\varphi\rangle/T\sim 1$ (this number can
be obtained in the lattice using a fully non-perturbative approach and
tracking Chern-Simons number in real time evolution). However this number
is very sensitive to $T$ and drops significantly when $T$ decreases.

Fully unsuppressed sphalerons can only be obtained in the global $U(1)$
case for large enough $\gamma/\lambda_S$. In fact, to obtain an asymmetry
of the order of the observed one, one would need $\gamma/\lambda_S\sim
10^{14}$. On the other hand, stability of the potential requires
$4\lambda/\gamma>\gamma/\lambda_S$, so that $\lambda/\lambda_S\sim
10^{28}$. Such an ad-hoc and wild fine-tuning of the parameters prevents
us from taking this particular case seriously.

Unsuppressed sphaleron transitions inside the string cores are not
sufficient to
generate the baryon asymmetry: they must occur in a background 
with CP asymmetric particle distributions so that the sign of the
B-violation is biased. This asymmetry comes about if the interactions
between the particles in the plasma and the string walls violate CP. In
that case the walls of a moving string act as sources of chiral-number
flux (which would be zero if the string velocity $v_S$ were zero).
This asymmetry diffuses away from the walls and only that inside the
string is useful to create baryons (for geometrical reasons it is also
clear that this diffusion effect is less efficient for strings than for
bubbles). In conclusion, we have to compute the chemical potential
$\mu$ for chiral number inside the strings. General arguments (confirmed
by detailed analysis of particular models) give the result $\mu=Kv_S^2T$
for small $v_S$, with $K\simlt 0.01$ and $\mu=K'T$ for $v_S\sim 1$ with
$K'$ of order 1.

\section{Evolution of string networks and efficiency of baryogenesis}

To get a final number for the asymmetry generated by this mechanism, we 
need to know how many strings there are and how quickly they are
moving (the best case being that of a dense network of fast moving
strings). We can describe the string network by a mean average separation
between strings $R(t)$ and a mean average velocity $v_S(t)$. The
evolution of these quantities with time $t$ is governed by Hubble
expansion ($H\sim 1/2t$); energy loss by loop formation; and friction
with
the plasma. The friction force goes like $F\sim v_ST^3$: it is
important at early times when it dominates the dynamics of the evolution.
This is the friction dominated or Kibble regime, with $R(t)\sim t^{5/4}$
and $v_S(t)\sim t^{1/4}\sim HR(t)$.
Eventually, friction will no longer be important and a scaling regime is
reached with $R(t)\sim 1/H$ and $v_S\sim 1$.

In conclusion, to get the final number for the baryon asymmetry we start
with the  equation for the rate of change of baryon number $N_B$ per unit
time and unit length of string:
\be
\label{eq}
\frac{dN_B}{dLdt}=1.5 [\kappa_l\alpha_w^2T^2]\frac{\mu}{T}.
\ee
If we use the results for $\kappa_l$ and $\mu$ previously discussed,
and integrate eq.(\ref{eq}) in one Hubble time (this is because
$\kappa_l$ is shut-off quickly with decreasing $T$) using the 
network evolution results just presented we end up with the result that
\be
\left[\frac{N_B}{N_\gamma}\right]_{strings}
\simlt
10^{-10}\left[\frac{N_B}{N_\gamma}\right]_{observed}.
\ee
That is, the mechanism just studied is uncapable of generating a
sufficiently large matter-antimatter asymmetry.

\section*{Acknowledgments}
I thank J.M. Cline, G.D. Moore and A. Riotto for an enjoyable
collaboration on the topic presented.

\end{document}